\documentclass{jpconf}

\usepackage{graphicx}
 \usepackage{amsmath,amssymb}
 \usepackage{xcolor}
\newcommand{\figurerightsidecaption}[5]{
  \begin{minipage}[t]{#1}
    \vspace*{0pt}
    \includegraphics[width=1.\textwidth]{#2}
  \end{minipage}
  \hfill%
  \begin{minipage}[t]{#3}
    \vspace*{0pt}
    \caption{#4}
    \label{#5}
\end{minipage}%
}

\newcommand{\be}{\begin{equation}}
\newcommand{\ee}{\end{equation}}
\newcommand{\bdm}{\begin{displaymath}}
\newcommand{\edm}{\end{displaymath}}
\newcommand{\bea}{\begin{eqnarray}}
\newcommand{\eea}{\end{eqnarray}}
\newcommand{\ba}{\begin{align}}
\newcommand{\ea}{\end{align}}

\newcommand{\PD}[2]{\partial_{#2} #1}
\newcommand{\TD}[2]{\frac{\mathrm{d} #1}{\mathrm{d}#2} }
\renewcommand{\vec}[1]{{\mbox{\boldmath${\mathrm{#1}}$} }}  
\newcommand{\DIV}[1]{\nabla\cdot {#1}}
\newcommand{\CURL}[1]{\nabla \times {#1}}

\newcommand{\bB}{\vec{B}} 
\newcommand{\bE}{\vec{E}} 
\newcommand{\bv}{\vec{v}} 
\newcommand{\bJ}{\vec{J}} 
\newcommand{\bey}{\vec{\hat{y}}}
\newcommand{\bez}{\vec{\hat{z}}}

\newcommand{\eg}{e.g.}
\newcommand{\ie}{i.e.}

\graphicspath{{./}}
 
\begin{document}

\title{Fast magnetic reconnection: The \textit{ideal} tearing instability
in classic, Hall, and relativistic plasmas.}

\author{E. Papini$^1$, S. Landi$^{1,2}$ and L. Del Zanna$^{1,2,3}$}

\address{$^1$Dipartimento di Fisica e Astronomia, Universit\`a degli Studi di Firenze, Italy.}

\address{$^2$INAF - Osservatorio Astrofisico di Arcetri, Firenze, Italy.}

\address{$^3$INFN - Sezione di Firenze, Italy.}

\ead{papini@arcetri.astro.it}

\begin{abstract}
Magnetic reconnection is believed to be the driver of many explosive phenomena in Astrophysics, from solar to gamma-ray flares in magnetars and in the Crab nebula. However, reconnection rates from classic MHD models are far too slow to explain such observations. 
Recently, it was realized that when a current sheet gets sufficiently thin, the reconnection rate of the tearing instability becomes ``ideal'', in the sense that the current sheet destabilizes on the ``macroscopic'' Alfv\'enic timescales, regardless of the Lundquist number of the plasma. Here we present 2D compressible MHD simulations in the classical, Hall, and relativistic regimes. In particular, 
the onset of secondary tearing instabilities is investigated within Hall-MHD for the first time.  
In the frame of relativistic MHD, we summarize the main results from Del Zanna et al.~\cite{2016delzannaMNRAS}: the relativistic tearing instability is found to be extremely fast, with  reconnection rates of the order of the inverse of the light crossing time, as required to explain the high-energy explosive phenomena.

\end{abstract}

\section{Introduction: Fast magnetic reconnection and the \textit{ideal} tearing instability}

Magnetic reconnection is believed to be the driver of many explosive phenomena observed in laboratory and astrophysical plasmas, such as  sawtooth crashes in tokamaks and solar flares, which involve the conversion of magnetic energy into heat and particle acceleration in the surrounding plasma.
Reconnection may also explain several high-energy events observed in extreme environments, for instance the gamma-ray flares observed in the Crab nebula and the soft gamma repeaters observed in magnetars, to name a few.
All these phenomena are impulsive, characterized by a violent release of energy, and happen on very short timescales.
However, classic models of reconnection predict characteristic times which, 
for highly conducting plasmas, are too slow to explain the rapid evolution of the observed events.
In the following we briefly review these models
and highlight recent breakthroughs from fast magnetic reconnection models in classic and relativistic plasmas.

In the frame of resistive magnetohydrodynamics (MHD)
the evolution of the magnetic field $\bB$ is governed by the induction equation
\begin{gather}
 \PD{\bB}{t} = \CURL{\left(\bv\times\bB\right)} +\eta\nabla^2\bB
\label{eq:induction}
 \end{gather}
where $\bv$ is the plasma flow velocity and $\eta$ is a (constant) magnetic diffusivity.
We remind that magnetic reconnection can occur only in presence of a non-zero resistivity (\ie~a finite value of~$S$).
From the above equation it is possible to identify two characteristic timescales, $\tau_A={L}/{c_A}$ and $\tau_D = {L^2}/{\eta}$, which describe the fluid advection and the Ohmic diffusion of magnetic fields respectively.
Here $c_A$ is the Alfv\'en speed and $L$ is a characteristic macroscopic length.
The ratio of the two timescales defines the Lundquist number $S={L c_A}/{\eta}$.
For astrophysical plasmas, $S$ is very high ($\sim 10^{12}$), thus reconnection is allowed only in current sheets with strong magnetic field gradients.

Early models of magnetic reconnection concern two-dimensional stationary equilibria. In particular, the Sweet-Parker model \cite{1958sweet,1957parker} (hereafter SP) of steady incompressible reconnection driven by velocity inflows $v_{in}$ predicts, for a current sheet of length $L$ and thickness $a$, a reconnection time 
\begin{gather}
 \tau_{SP}  = \tau_A ({c_A}/{v_\mathrm{in}}) = \tau_A ({L}/{a}) =  \tau_A S^{1/2}
\end{gather}
which increases with $S$, thus far too slow to explain the observed events.
Note that the SP current sheet must have a fixed aspect ratio $L/a$ scaling as $S^{1/2}$.
Concurrently,  Furth et al. \cite{1963FKR} found that an infinitely long currentsheet is locally prone to the  linear \textit{tearing} instability, which leads to the formation of multiple X-points and magnetic islands (plasmoids) during the reconnection process. The $e$-folding growth time $\tau_t$ of the fastest reconnecting mode, calculated by using the current sheet thickness $a$ as characteristic length scale,   
reads 
\begin{gather}
 \tau_t  \simeq \tau_A^aS_a^{1/2}\,/\,0.6,\quad S_a={a c_A}/{\eta}
 ,\quad  \tau_A^a = {a}/{c_A },
 \label{eq:tearing_gamma}
\end{gather}
again too slow and increasing with the Lundquist number $S_a$.
However, by considering the onset of the tearing instability in extremely thin current sheets and renormalizing all quantities in Eq. (\ref{eq:tearing_gamma}) with respect to the global scales $L$, one obtains linear growth rates $\gamma=1/\tau_t$ which are very fast and increase with $S$ for thin enough current sheets.
In particular, recent theoretical and numerical works (see, \eg,~ \cite{2007loureiro, 2008lapenta, 2009samtaney, 2009bhattacha, 2009cassak}) have shown that, for high $S \gtrsim 10^4 $ and for a current sheet with a Sweet-Parker aspect ratio $L/a\sim S^{1/2}$, reconnection is indeed very fast, $\gamma$ increasing with $S$ as
$\gamma \tau_A \sim S^{1/4}$, and leads to the formation of a chain of plasmoids. 

However, the existence of this plasmoid instability brings to a paradox, since it predicts an infinite reconnection rate as  $S \rightarrow \infty$, even though reconnection is impossible in ideal MHD.
This paradox can be avoided if we proove that SP current sheets cannot possibly form. 
In fact, the plasmoid instability is so fast that a current sheet could be disrupted by the tearing instability before reaching the desired aspect ratio of SP.
To assess this possibility, it is convenient to analyze the mechanisms at the base of current sheet formation.

We can identify four  mechanisms at play into the dynamic evolution of a forming current sheet: the diffusion of the bulk magnetic field, the pile-up of plasma toward the current sheet center due to the presence of inflows with velocity $v_{in}$, the expulsion of plasma along the current sheet outflows, and finally the tearing instability, which tend to disrupt the current sheet.
Accordingly, the characteristic time scales associated to the above mechanisms are
\begin{gather}
 \tau_D = {a^2}/{\eta},\quad 
 \tau_{up} = {a}/{v_{in}} = \tau_A ({a}/{L}) ({c_A}/{v_{in}}),\quad
 \tau_A={L}/{c_A},\quad
 \tau_t =\tau_A S^{1/2} \left({a}/{L}\right)^{3/2}
\end{gather} 
where $\tau_t$ has been renormalized with respect to $L$ and we assumed the outflow to be  alfv\'enic, that is a low limit for the expulsion time. 
The tearing instability is always faster than diffusion in the cases where $S > L/a$, which is true for SP current sheets, but also for sheets with a smaller aspect ratio.
The presence of an outflow has a stabilizing effect, since plasmoids can be expelled before they grow enough to destabilize the current sheet. 
The plasma pile-up is the main formation mechanism for the current sheet.
Hence the current sheet disrupt if the conditions 
\begin{gather}
  {\tau_t} < {\tau_A} \, \Longleftrightarrow\, {a}/{L} < S^{-1/3} 
  ,\quad \text{and }\quad
  \tau_t < \tau_{up} \, \Longleftrightarrow \, {a}/{L} < S^{-1} 
  \left ({c_A}/{v_{in}}\right )^2
  \label{eq:tt_tup}
 \end{gather}
are satisfied.
The first condition fixes the critical aspect ratio at which the current sheet is disrupted
\begin{gather}
 {L}/a = S^{1/3},
 \label{eq:ideal_aspect_ratio}
\end{gather}
which already implies that a SP current sheet will never form, since it would be destroyed by the tearing instability before reaching the ratio $L/a=S^{1/2}$.
The further matching of the second condition (\ref{eq:tt_tup})
sets the scaling for the inflows, namely
$v_{in}/c_A = S^{-1/3}.$
A current sheet with a critical aspect ratio has another important property: the growth rate of the fastest tearing mode becomes \textit{ideal}, in the sense that is independent on the Lundquist number and of the order of the macroscopic Alfv\'en time, \ie,  $\tau_t\sim\tau_A$.
The ideal tearing instability was studied theoretically by \cite{2014puccivelli}, who calculated the tearing eigenmodes in a current sheet of ideal aspect ratio $L/a=S^{1/3}$ and found that the linear growth rate of the fastest reconnecting mode asymptotically tends to $\gamma\tau_A \simeq 0.67$, in agreement with Eq.~(\ref{eq:tearing_gamma}).
The results have been  confirmed  and extended to the nonlinear regime by \cite{2015landi} and \cite{2016delzanna} using numerical simulations.
It is worth noting that, as found by \cite{2016uzdensky}, if we consider a more general non-alfv\'enic outflow $v_\text{out}$, Eq. (\ref{eq:ideal_aspect_ratio}) then modifies to $L/a = S^{1/3} (v_\text{out}/c_A)^{2/3}$, granted that the current sheet is sufficiently thin to host the most unstable tearing eigenmode. Also in this case, \cite{2016uzdensky} demonstrated that it is not possible to form a SP current sheet.

In the following sections we further investigate the nonlinear regime of the fast magnetic reconnection in the MHD case, and extend the study to include the effect of Hall currents, which may be relevant at reconnection sites where the characteristic scales of the tearing instability become comparable to the ion inertial length. 
Finally, we will present the results of a recent work \cite{2016delzannaMNRAS} which generalizes the ideal tearing instability to include relativistic regimes.

\section{Including the Hall term in the \textit{ideal} tearing instability}

We first discuss the tearing instability in the case of weakly collisional plasmas. 
At spatial ion scales, the electron and the ion velocities decouple. 
Since the electrons rather than the ions drive the magnetic field, 
the velocity term in the induction equation (\ref{eq:induction}) is replaced by the electron velocity
 $\bv_e = \bv - \bJ/(en_e) $,
where $n_e$ is the numerical density of electrons, $e$ is the unsigned fundamental electrical charge, and $\bJ = c/(4\pi)\nabla \times \bB$ is the current density. 
Neglecting the effects of the electron inertia and of the electron pressure,
there is only one new term in the induction equation, with respect to its MHD counterpart, that is the Hall term:
\begin{align}
 -\nabla\times \left [ \frac{\bJ}{en} \times \bB \right] =
 -\nabla\times\left [ \frac{c m_i}{4\pi e \rho} (\nabla\times\bB) \times \bB\right ],
 \label{eq:hall_term}
\end{align}
where we used charge neutrality $n_e = n_i=n$ and approximated the plasma density with the ion density $\rho = m_i n$.
To understand how the Hall term changes the behavior of the standard MHD tearing instability we need to identify the characteristic scale at which it becomes important.
The fundamental parameter is the ion inertial scale 
 $d_i = {c}/{\omega_{pi}} = c \sqrt{{m_i}/({4\pi n_i e^2})}$,
which depends on the ion plasma frequency $\omega_{pi}$.
We renormalize the induction equation using the Alfv\'en speed $c_A = {B_0}/{\sqrt{4\pi\rho_0}}$ and time  $\tau_A = {L}/{c_A}$, where $B_0$ and $\rho_0$ are the asymptotic values of the magnetic field and the density respectively, far from the current sheet, and $L$ is the length of the current sheet.
We then obtain 
\begin{align}
 \PD{\vec{B}}{t} = \nabla\times\left ( \bv \times \bB\right ) + \frac{1}{S} \nabla^2 \bB
 -  \eta_H\nabla\times\left ( \frac{1}{\rho} (\nabla\times\bB) \times \bB\right )
  \label{eq:induction_hall_adi}
\end{align}
where all quantities and operators are now renormalized with respect to $L,\tau_a,B_0$, and $\rho_0$,  
and we introduced the Hall resistivity
\begin{equation}
 \eta_H = \frac{d_{i0}}{L} = \frac{c m_i}{L\sqrt{4\pi e^2 \rho_0}}
\end{equation}
with $d_{i0}$ being the asymptotic value of the ion inertial length.
The Hall term is not negligible when the ion inertial length becomes comparable to the width $\delta$ of the inner resistive layer of the tearing instability \cite{2017puccivelli}. 
For the fastest growing mode, the inner width $\delta$ is described by the equation 
 ${\delta}/{a} \simeq S_a^{-3/10} \Delta'^{1/5}$\cite{1993biskamp},
where $S_a$ is given in Eq. (\ref{eq:tearing_gamma}), and $\Delta'$ is an instability parameter which depend on the configuration considered for the equilibrium magnetic field.  $\Delta'$ may depend on $ka$, which for the fastest growing mode scales as $ka\sim S_a^{1/4}$. For instance, in the equilibrium configuration considered 
by \cite{2007loureiro} $\Delta'$ is indipendent of $S_a$, while for the Harris sheet configuration considered in this work  (see next Section) 
 $\Delta' = 2 \left ( {1}/{ka} - ka \right )$.
Thus, at high Lundquist numbers $S_a \gg 1$ and for the fastest growing mode, $ka\sim S_a^{1/4}$, we have $ \Delta' \simeq 2 S_a^{1/4}$, and $\delta/a \simeq S_a^{-1/4}$.
Therefore, by rescaling to the macroscopic length $L$, the ratio  
\begin{align}
{d_{i0}}/{\delta} =  \eta_H \left ({a}/{L}\right )^{-3/4} S^{1/4}
\end{align}
determines whether Hall effects are important in the dynamics of reconnection.
For a current sheeet with an aspect ratio $L/a =S^{\alpha}$ that scales with a power of $S$, one obtains
 ${d_{i0}}/{\delta} = \eta_H  S^{(3\alpha+1)/4}$.
Finally, by considering a current sheet of ideal aspect ratio ($\alpha=1/3$), we obtain
\begin{align}
 {d_{i0}}/{\delta} =  \eta_H  S^{1/2}.
\end{align}
We can identify three regimes: a MHD regime ($ \eta_H \ll S^{-1/2}$) where the Hall term does not play a relevant role, a mild Hall regime ($\eta_H \lesssim S^{-1/2}$), where the ion inertial scale is comparable to the thickness of the inner layer,
and a strong Hall regime ($\eta_H \gg S^{-1/2}$), where reconnection is dominated by the Hall effect and the classic theory of the tearing instability is no longer valid.   
The linear phase of the tearing instability in these three regimes has already been investigated by \cite{2017puccivelli} in the case of a Harris current sheet in pressure equilibrium. 
They verified the existence of these regimes, and showed that the linear growth rate start increasing for, \eg, values $d_i/\delta \sim 3$ at $S=10^6$.   
However this threshold is likely smaller, since Hall currents may affect the subsequent nonlinear evolution, where thinner current sheets formed between the ejected plasmoids may eventually host secondary reconnection events, as we will show in the next section.

\section{Hall effects in the nonlinear evolution of magnetic reconnection}
\subsection{Numerical setup}
Building on our previous works \cite{2015landi,2016delzanna,2017landi}, we investigate the nonlinear evolution of a reconnecting current sheet, by means of 2D resistive Hall-MHD simulations.
Togheter with the induction equation (\ref{eq:induction_hall_adi}), we integrate numerically the compressible Hall-MHD equations in the form:
\begin{gather}
 \PD{\rho}{t} + \DIV{(\rho \bv)} = 0\,,
  \label{eq:continuity}
  \\
 \rho\left (\PD{}{t} + \bv\cdot\nabla \right) \bv = -\nabla P + (\CURL{\bB})\times\bB\,,
  \\
  \left (\PD{}{t} + \bv\cdot\nabla \right) T = 
  (\Gamma -1 ) \left [ - (\DIV{\bv})T 
  + \frac{1}{S} \frac{|\CURL\bB|^2}{\rho} \right ]
\end{gather}
where $\Gamma=5/3$ is the adiabatic index and all quantities retain their obvious meaning. Velocities, lengths, and time are expressed in units of the Alfv\'en speed $c_A$, the current sheet length $L$, and the Alfv\'en time $\tau_A$ respectively. The pressure is normalized to $B_0^2/4\pi$ and the energy equation is written in terms of the normalized temperature $T=P/\rho$ and expressed in units of $T_0 = (m_i/k_B) c_A^2$.
With the chosen normalization the Lundquist number is the inverse of the magnetic diffusivity $S=\eta^{-1}$. 
For the Lundquist number we use $S=10^6$ as our reference value.
We focus on a initial force-free equilibrium and consider a Harris current sheet
\begin{gather}
 \bB = B_0 \tanh(x/a) \bey + B_0 \,\mathrm{sech} (x/a) \bez,
 \label{eq:harris_ff}
\end{gather}
with homogeneous density $\rho_0=1$ and pressure $P_0=\beta_0/2$, and with no velocities $\bv_0=0$. 
The asymptotic field magnitude is $B_0=1$ and the plasma beta is set to $\beta_0=0.5$.   
The thickness of the current sheet is set to have an ideal aspect ratio 
$a/L = S^{-1/3}=\eta^{1/3}$.
As it is defined, the current sheet is centered at $x=0$ and is directed along the $y$-direction.

We consider a rectangular domain of size $[-L_x,L_x]\times [0,L_y]$.
The value of $L_y$ is chosen such that the lowest wavenumber resolved for the tearing instability is $ka = 2\pi a/L_y = 0.02 $. This is more than sufficient to resolve the fastest growing mode of the tearing instability for $S=10^6$.
In the $x$-direction we set $L_x=20a$ in order to have sufficiently far boundaries from the reconnection region while retaining the high resolution required inside the current sheet with a reasonable amount of grid points.   

The Hall-MHD equations are numerically solved by means of the same MHD code we used in \cite{2015landi}, modified to include the Hall term.
We impose periodic boundary conditions in the $y$-direction 
and open boundary conditions in the $x$-direction using the method of projected characteristics.
Spatial derivatives are calculated using Fourier decomposition along the periodic direction and a fourth-order compact scheme \cite{1992lele} across the current sheet. Time integration is performed with a third-order Runge-Kutta scheme. For further details we refer to \cite{2005landi}.

The grid employed consists of $N_x \times N_y = 4096\times512$ points, which allows to resolve secondary reconnection events in both the $x$- and $y$-directions.

\subsection{Results}
We run several numerical simulations employing different values of the Lundquist number ranging from $S = 10^5$ to $10^7$ and for different values of the Hall diffusivity.
To trigger the tearing instability, the initial equilbrium was perturbed  by sinusoidal magnetic perturbations of amplitude $\epsilon \sim 10^{-4}$, localized in the current sheet.
For the perturbations we used the same analytical form of \cite{2015landi} but for the magnetic field instead of the velocity.
We excited the first ten modes with $ka$ ranging from $0.02$ to $0.2$.

\begin{figure}[t]
\figurerightsidecaption{0.555\textwidth}{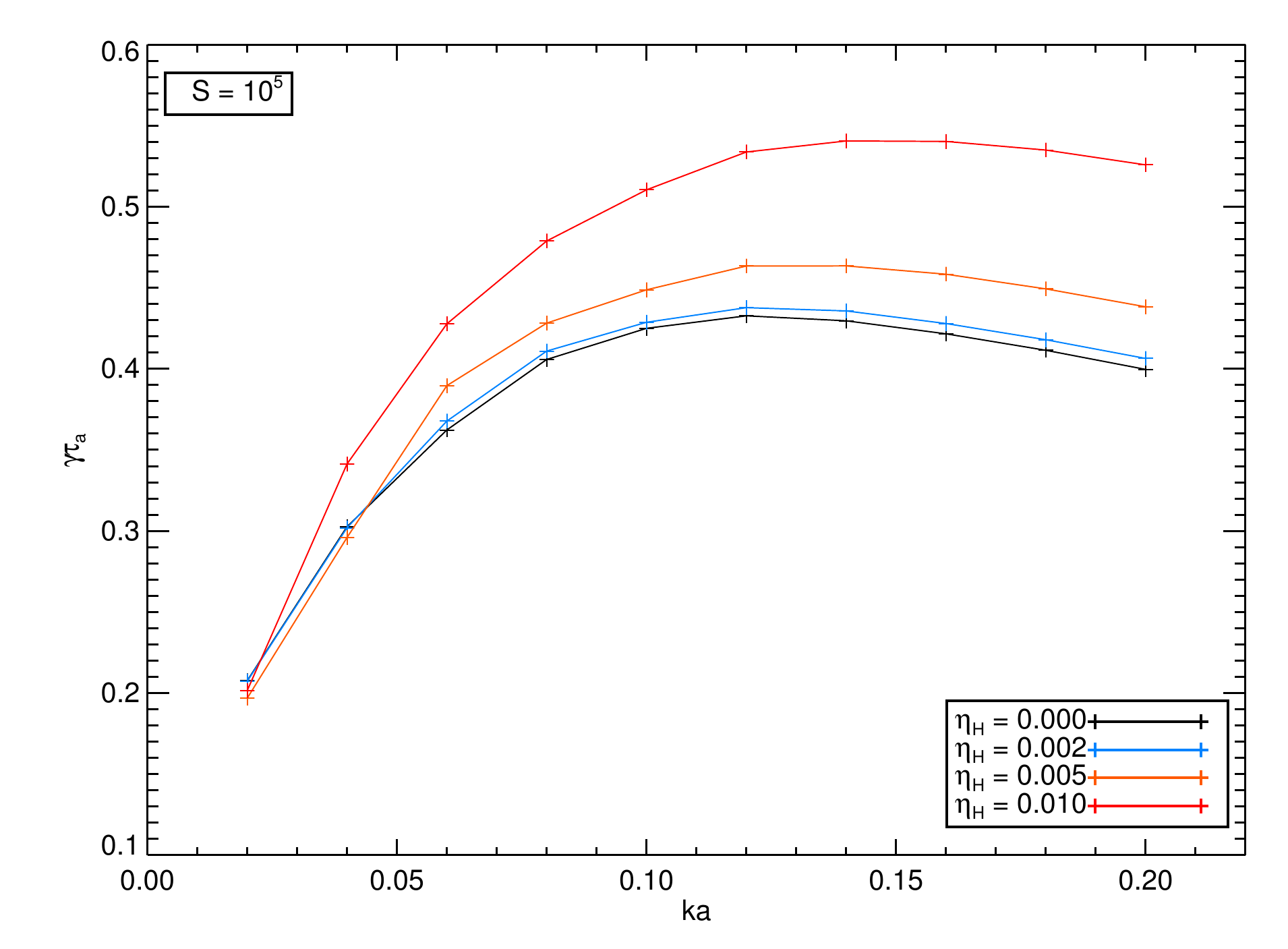}{0.42\textwidth}{\begin{footnotesize}Growth rates $\gamma\tau_A$ vs. $ka$ in the linear phase of Hall-MHD simulations with $S=10^5$ and different values of $\eta_H$.
 The first ten wavenumbers with $ka$ from $0.02$ to $0.2$ have been excited.
 \end{footnotesize}}{fig:gammatauvska10e5}
\end{figure}

In all the simulations, a linear tearing instability develops at the beginning and with the same qualitative behavior, but with quantitatives differences. 
As an example, Fig.~\ref{fig:gammatauvska10e5} shows the dispersion relation calculated in the linear phase for four simulations with the same value of $S=10^5$ and different values of $\eta_H=0$ (MHD case), $ 0.002, 0.005,\text{ and } 0.01$, corresponding to $d_i/\delta = 0, 0.6, 1.6, \text{and } 3.2$ respectively.
The dispersion relation has the same shape in all cases. 	
However, the linear growth rate of each mode increases  when $d_i$ exceeds the thickness of the inner resistive layer $\delta$, up to about $25\%$ more than the MHD case for $ka=0.14$ and $\eta_H=0.01$. This is in qualitative and quantitative agreement with \cite{2017puccivelli} even though the initial equilibrium here is different and therefore the linear evolution may also be different, due to the additional terms present in the Hall-MHD linearized equations.

\begin{figure}
 \includegraphics[width=\columnwidth]{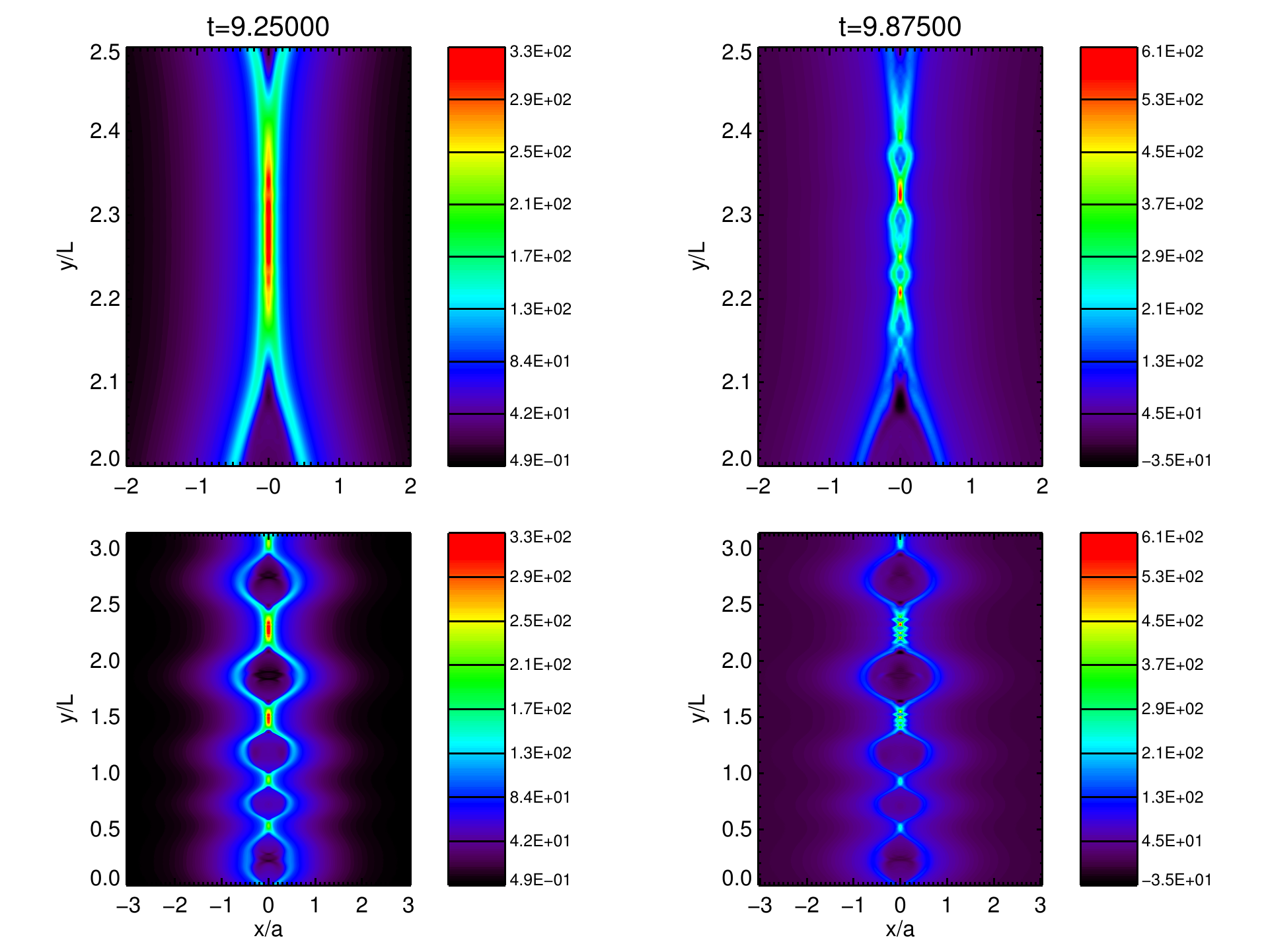}
 \caption{\begin{footnotesize}
Filled Contour of $J_z$ in the MHD case with $S=10^6$ and at two different times, before a tearing instability is triggered in the secondary current sheets (left), and at the time when the secondary reconnection events reach the nonlinear phase (right). Bottom panels show $J_z$ for the full length of the current sheet, top panels zoom into one of the secondary sheets. Note that $x$ and $y$ are normalized on different scales.
\end{footnotesize}}
 \label{fig:jzcontour}
\end{figure}
At the beginning of the nonlinear phase, that is at $t=9.25~\tau_A$ in our reference run (bottom left panel of Fig.~\ref{fig:jzcontour}), the size of the plasmoids is comparable to the current sheet thickness, and some of them already merged. Between plasmoids secondary current sheets have been formed, with a thickness of roughly one tenth of the original thickness. One of them is shown on the top left panel of Fig.~\ref{fig:jzcontour}.
The subsequent evolution is characterized by the coalescence and nonlinear growth of plasmoids, but the most important feature is the onset of secondary reconnection events in the newly formed current sheets, which then drive the dynamics and eventually lead to the disruption of the whole system.
These secondary tearing instabilities are indeed very fast, since at time $t=9.875~\tau_A$ (right panels of Fig.~\ref{fig:jzcontour}) they already fully developed, in less than $0.5$ (macroscopic) Alfv\'en times.

\begin{figure}
 \includegraphics[width=0.5\textwidth]{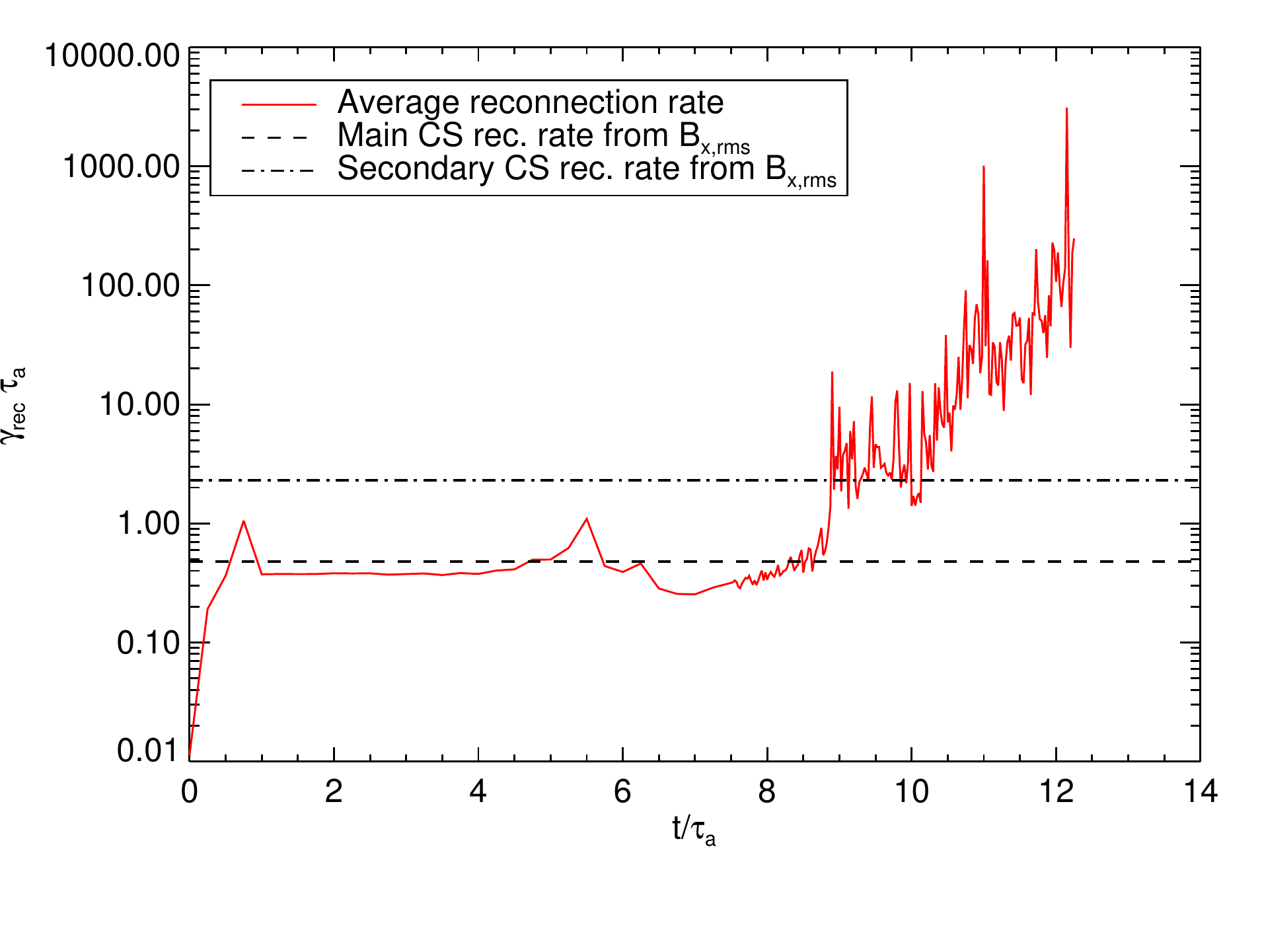}
 \includegraphics[width=0.5\textwidth]{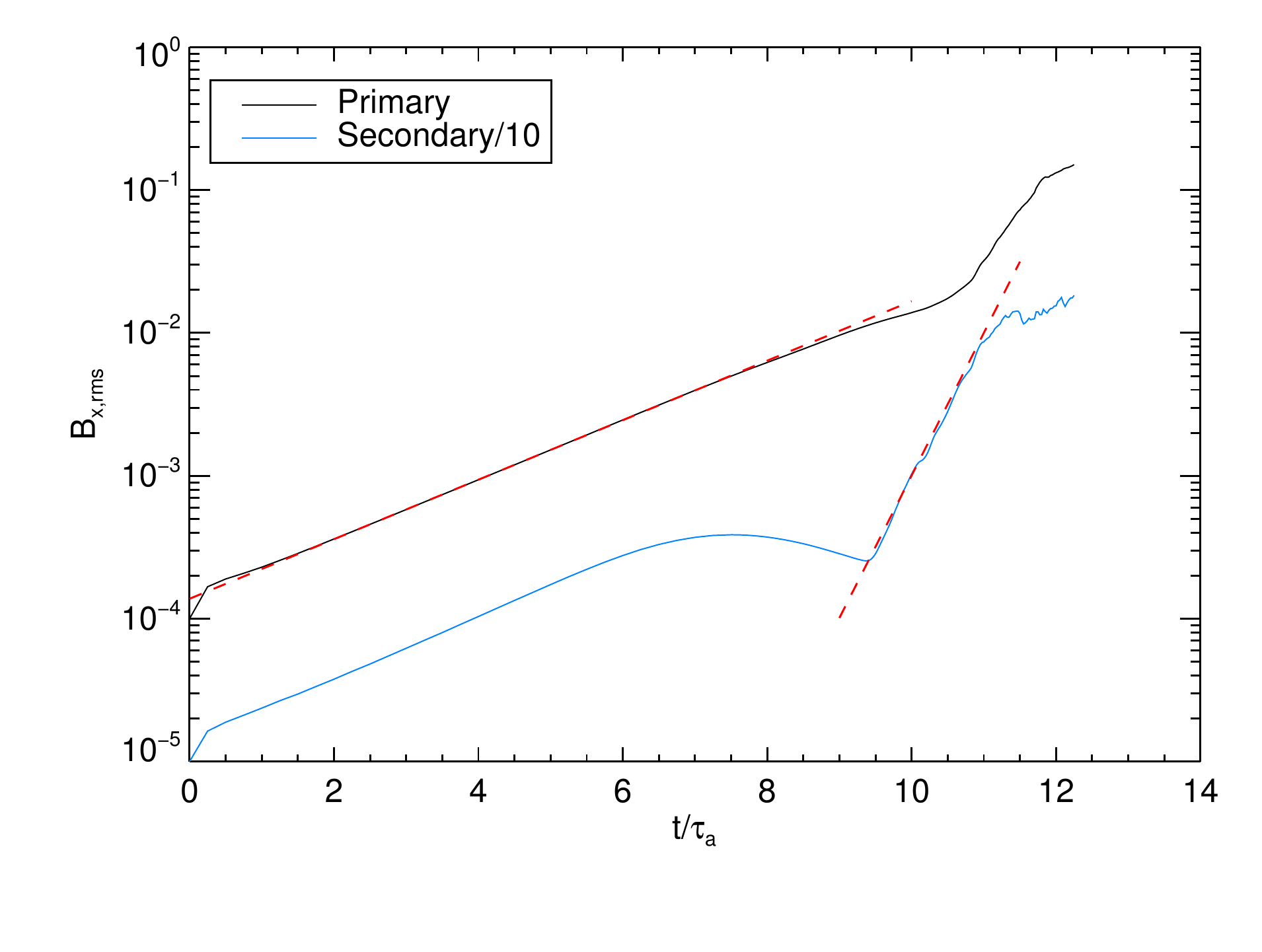}
 \caption{\begin{footnotesize}\textbf{Left:} Averaged reconnection rate $\gamma_\mathrm{rec}\tau_A$ vs. time for the MHD case $S=10^6$.
 Horizontal lines are the linear growth rate as calculated by an exponential fit of $B_{x,\mathrm{rms}}$ (right panel).
 \textbf{Right}: $B_{x,\mathrm{rms}}$ vs. time of the primary current sheet (black solid line) and rms divided by ten of the secondary sheet of top panels of Fig.~\ref{fig:jzcontour} (blue solid line). Red dashed lines denoted the exponential fit for the growth rates of the primary and secondary reconnection events.
 \end{footnotesize}}
 \label{fig:recrates1e6h0}
\end{figure}
To quantitatively support the above statements, we calculated the averaged reconnection rate $\gamma_\mathrm{rec}$ of the current sheet,
shown in the left panel of Fig.~\ref{fig:recrates1e6h0} for the MHD reference run.
The averaged reconnection rate 
\begin{gather}
 \gamma_\mathrm{rec} = \frac{1}{N}\sum_{i=1}^N \frac{1}{\Phi_i}\TD{\Phi_i}{t},
\end{gather}
is calculated by taking, at each time, the logarithmic time derivative of the reconnected flux $\Phi_i$ between the $i$-th pair of X- and O-points and then averaging over all the pairs, $N$ being the total number of pairs in the current sheet.
In Fig.~\ref{fig:recrates1e6h0}, after the initial perturbations reorganized to select the tearing modes, a clear plateau is present up to $t\simeq 9 \tau_A$ and with a value $\gamma_\mathrm{rec} \tau_A\simeq 0.4$. In the same plot it is also possible to identify a second, more noisy plateau, corresponding to the average reconnection rate of the secondary current sheets. To better estimate the reconnection rates we performed a logarithmic fit of the rms of the $x$-component of the magnetic field $B_{x,\text{rms}}$, shown in the right panel of Fig.~\ref{fig:recrates1e6h0}
for $B_{x,\text{rms}}$ calculated over the whole current sheet (black line) and by restricting to the region of the secondary current sheet in top panels of Fig. \ref{fig:jzcontour} (blue line).
In the latter we clearly see the onset of the secondary tearing instability at around $t\simeq 9.5 \tau_A$. The horizontal dashed and dot dashed lines in the left panel, corresponding to the values $\gamma_\mathrm{rec}\tau_A = 0.48$ and $2.30$ given by the two fits, nicely match the two plateaux we identified. 
Indeed, the measured reconnection rate of the secondary current sheet is strongly superalfvenic and consistent with the preliminary estimate.
In the final stage of the evolution (not shown here) the secondary reconnection events drive the dynamics, with new plasmoids being ejected by superalfvenic outflows and feeding the huge plasmoids generated by the first reconnection event. Eventually, the entire current sheet is disrupted.

\begin{figure}[t]
\figurerightsidecaption{19pc}{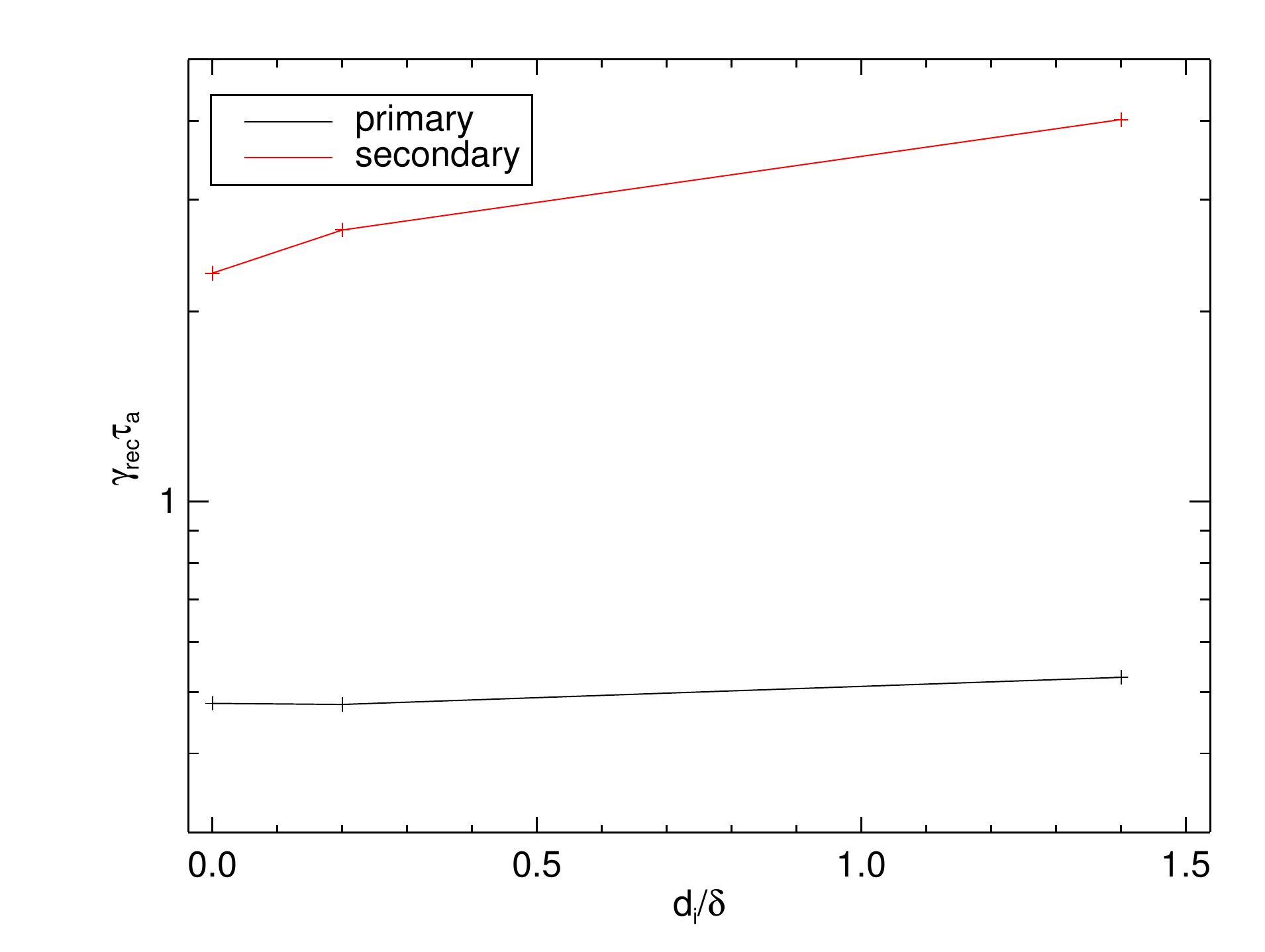}{17pc}{\begin{footnotesize}Reconnection rates, calculated with an exponential fit of $B_{x,\mathrm{rms}}$, of primary (black) and secondary (red) current sheets, for three simulations with  $S=10^6$ and $d_i/\delta=0, 0.2$, and $1.4$.\end{footnotesize}}{fig:recrates_hall}
\end{figure}
Even though for values of $d_i/\delta \lesssim 1$ the Hall term does not affect the reconnection rates of the primary current sheet, as we already shown in Fig.~\ref{fig:gammatauvska10e5}, the changes in the secondary reconnection events are drammatic.
Figure~\ref{fig:recrates_hall} shows a plot of $\gamma_\mathrm{rec}$ for the primary and the secondary reconnection events for three simulations with  $S=10^6$ and different values of the ratio $d_i/\delta$.
There, the reconnection rate of the secondary events (red lines) increases by $17\%$ for a value $d_i/\delta=0.01$, and almost doubles for $d_i/\delta=1.4$.
As a consequence, the evolution of the overall reconnection process is faster, and leads to the disruption of the current sheet in a shorter time.

Going to larger ratios $d_i/\delta$ we enter in the strong Hall regime: the primary reconnection events becomes more and more violent, and the formation of secondary current sheets seems to be inhibited.

\section{Ideal tearing instability in relativistic plasmas}
What we presented so far applies to classical plasmas.  
In what follows we review the main results of our recent work \cite{2016delzannaMNRAS}, which extend the study of the ideal tearing instability to plasmas in extreme astrophysical environments. There, magnetic fields are so strong that the Alfv\'en speed approaches the speed of light and the plasma becomes relativistic. It is convenient to introduce the plasma magnetization $\sigma_0 = B_0^2/\rho_0$ and the enthalpy density $w_0=\rho_0+4P_0$,
where we let $c\rightarrow 1$ and $4\pi\rightarrow 1$. With this normalization the plasma beta is $\beta_0=2P_0/B_0$.
At relativistic energies, the Alfv\'en speed takes the form
 $c_A={B_0}/\sqrt{w_0 +B_0^2}= ({1/\sigma_0 +2\beta_0 +1})^{-1/2}$,
that verifies $c_A < c$. Here the enhanced inertia is due to all the energetic contributions, as it must be in a relativistic case.

To study the relativistic tearing instability we use the same equilibrium configuration (\ref{eq:harris_ff}) of classical and Hall MHD with no equilibrium velocities and no electric fields.
We then introduce perturbations with small non-relativistic velocities
($v_1 \ll 1$). With the above assumptions, the linearized induction equation has the same form as in the MHD case, while the linearized momentum equation reads
\begin{gather}
 \PD{\left ( w_0\bv_1 + \bE_1\times \bB_0 \right)}{t} =
 - \nabla (P_1 + \bB_0\cdot \bB_1) +\bB_0\cdot\nabla\bB_1
 + \bB_1\cdot\nabla\bB_0
\end{gather}
where the subscripts zero and one denote equilibrium and perturbed quantities respectively. By assuming $\bE_1  \times \bB_0 \simeq B_0^2\bv_1$ (which was numerically verified), we obtain
\begin{gather}
 \PD{\left ( w_0\bv_1 + \bE_1\times \bB_0 \right)}{t} \simeq (w_0 + B_0^2)\PD{\bv_1}{t}  
\end{gather}
and the linearized equations for the tearing instability take the same form of the MHD case, provided all quantities are normalized with respect to the enhanced inertia $w_0 +B_0^2$ instead of $\rho_0$.
The linear analysis then proceeds as in the classic case, and we obtain the linear growth rate of Eq.~(\ref{eq:tearing_gamma}) for the relativistic tearing instability, and the growth rate $\gamma\tau_A \simeq 0.67$ for a current sheet of ideal aspect ratio $L/a = S^{1/3}$.
Differences between the relativistic and the classic tearing mode become evident if we renormalize the time scales with respect to the light time $\tau_c=L/c$ instead of the Alfv\'en time. Then one obtains that the maximum reconnection rate for a relativistic ideal tearing is
\begin{gather}
 \gamma\tau_c \simeq 0.6 (c_A/c).
\end{gather}
This is a novel result, which suggests that the relativistic magnetic reconnection triggered by the tearing instability may occur on light crossing time scales in extremely thin current sheets when $c_A\rightarrow c$, compatible with the burstly events observed in many astrophysical objects.

\section{Numerical simulations of relativitic fast magnetic reconnection}

To corroborate our theoretical findings and to study the full evolution of relativistic magnetic reconnection, we performed numerical simulations by means of the Eulerian conservative High-Order (ECHO) code \cite{2007delzanna}, which solves the resistive relativistic MHD equations combining shock-capturing properties and accuracy.
In ECHO, time integration is performed through a 3rd-order, 4 stages Strong stability Preserving IMplicit-EXplicit Runge-Kutta (SSP3-IMEX-RK) scheme \cite{2014delzanna,2014bugli}, while spatial integration is performed with an MP5 limited reconstruction combined with the HLL two-wave Riemann solver. For further details on the ECHO code, see \cite{2007delzanna}.
For the setup we used the same initialization of the MHD case for both the equilbrium and the initial perturbations. 
The size of the rectangular domain is the same. However, here we employed a different grid size of $1024\times512$ cells.

\begin{figure}
\figurerightsidecaption{0.5\textwidth}{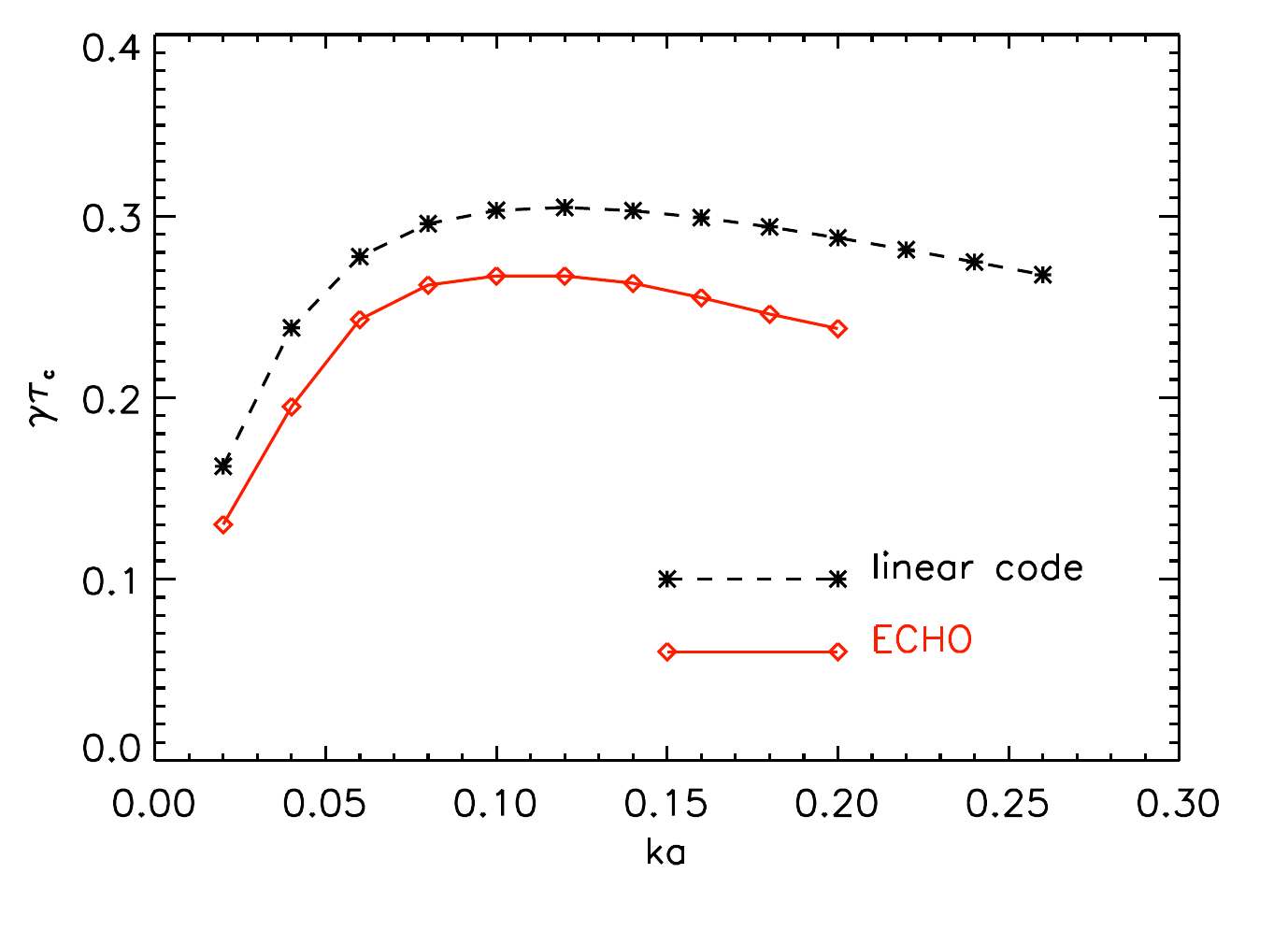}{.5\textwidth}
{\begin{footnotesize}
 Growth rates $\gamma\tau_c$ vs. $ka$ in the linear phase of the ideal relativistic tearing instability for a run with $\sigma_0 =1$, $\beta_0= 1$ (\ie, $c_A/c=0.5$), and $S=10^6$. 
 Red solid lines and diamonds are obtained with the ECHO code, while black dashed lines and asterisks are calculated by means of a classical linear MHD code were we set $\rho_0=4$ to obtain $c_A=0.5$ to mimic the linear relativistic behavior. The curve of the linear code exactly matches the theoretical dispersion relation for the ideal tearing instability.  
 \end{footnotesize}
}{fig:gammavska_relativistic}
\end{figure}
Figure~\ref{fig:gammavska_relativistic} shows the dispersion relation $\gamma(k)$, calculated with ECHO (red solid curve) by performing a single-mode analysis as done in \cite{2015landi}, for $S=10^6$ and $c_A=0.5$.
For comparison the same dispersion relation (black dashed curve) has been calculated by means of the linear pseudospectral MHD code employed in \cite{2015landi}, by setting for the latter $B_0=1$ and $\rho_0=4$ in order to obtain the same value for the Alfv\'en speed.
The curve from the linearized code exactly matches the theoretical prediction, with the fastest growing mode having $\gamma\tau_c=0.6c_A/c=0.3$. Here the discrepancy between the linear code and ECHO are likely due to the diffusion of the magnetic field as the simulation evolves.
Overall, the results confirm that the relativistic linear tearing modes grow independently of $S$ and on time scales of the order of the light crossing time. Additional numerical simulations performed going to highly magnetized plasmas ($\sigma_0 \gg 1$) and small beta values ($\beta_0 \ll 1$), have confirmed that the tearing instability gets faster as the Alfv\'en speed approaches the speed of light, and eventually occurs on time scales comparable to the global light crossing time.

To investigate the nonlinear regime of the relativistic tearing instability, we performed a set of five simulations for the same value of the Lundquist number $S=10^6$ but different values of plasma magnetization and plasma beta, in order to obtain for the Alfv\'en speed different values in the range from $0.5c$ up to $0.98c$. For these simulations we set $L_y=2L$.
In all simulations, the nonlinear phase is characterized by the growth and coalescence of the plasmoids, as in the classic MHD scenario.
The final evolution brings to the formation of a single and well defined X-point, where a secondary reconnection event occurred, and to the coalescence of all plasmoids into a single, large one.
Given the amplitude of the initial perturbations ($\sim10^{-4}$), the whole reconnection process last within $\sim 20 \tau_c$.
\begin{figure}
 \includegraphics[width=\textwidth]{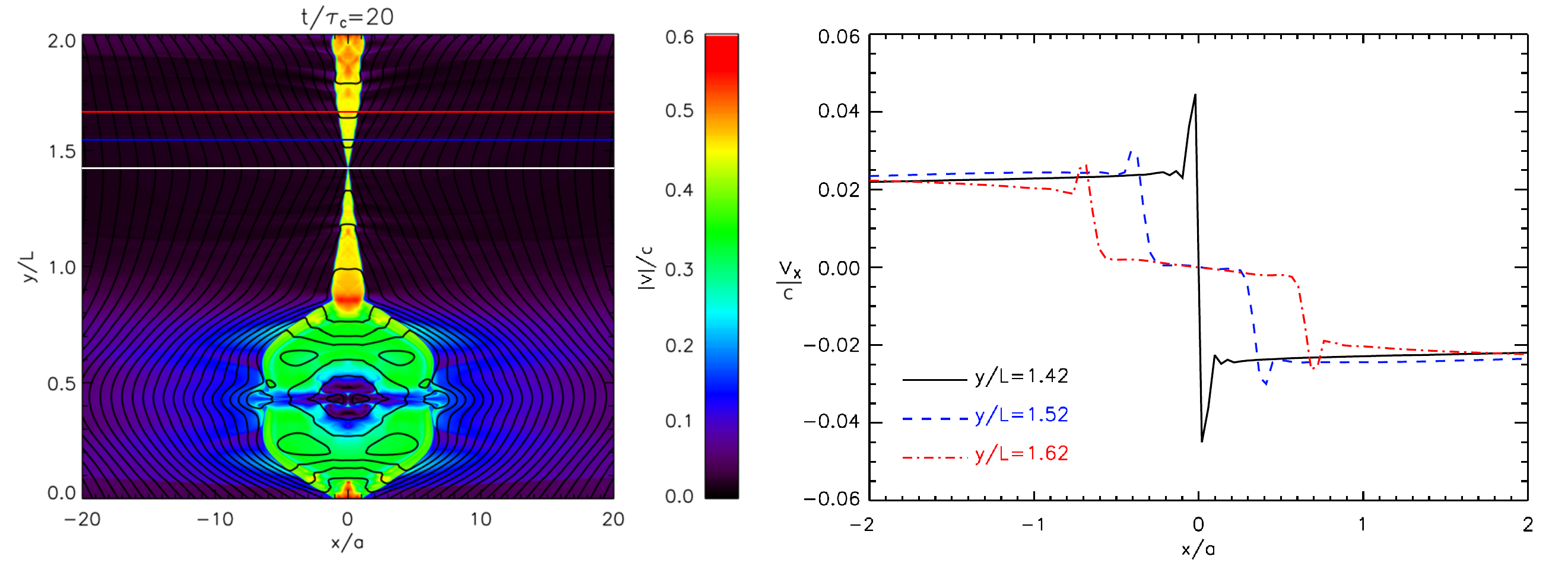}
 
 \caption{\begin{footnotesize}\textbf{Left}: color contour of the plasma velocity amplitude $|\bv|$ at $t=20\tau_c$. 
  Black curves denote magnetic field lines. Horizontal white, blue, and red lines indicate the location of the cuts of $v_x$ shown in the right panel.
 \textbf{Right}: cuts of $v_x$ across the current sheet, at the X-point and at two different locations along one of the jets.  
 \end{footnotesize}}
 \label{fig:petschek_v}
\end{figure}

The final state (shown in the left panel of Fig.~\ref{fig:petschek_v}) is most interesting, as it shows a Petschek-like configuration \cite{1964petschek}.
Here superalfvenic jets originating from the X-point are channeled into an exhaust and feed the plasmoid through a fast magnetosonic shock.
The jets are mildly relativistic, with a peak velocity of $\simeq 0.55 c$ at the termination shock. These results are in agreement with previous works (see,~\eg, \cite{2010zenitani, 2011zanotti}).
Additional simulations (not shown here) were performed by starting from a current sheet in pressure equilibrium. All confirm that the Petschek configuration is the natural final state of this tearing instability, independently on the initial equilibrium considered.

To compare our results with the Petschek model more quantitatively, we calculated the reconnection rate 
\begin{gather}
 \mathcal{R} \equiv M_A = {|v_x|}/{c_A} = {\pi}/({4\ln S})
\end{gather}
for a steady Petschek model. By using $S=10^6$, we predict a reconnection rate of $\mathcal{R}\simeq 0.057$.
Right panel of Fig.~\ref{fig:petschek_v} shows a roughly constant inflow velocity $|v_x|\simeq 0.025$ at different distances from the X-point in the exhaust, which gives a value of $M_A\simeq 0.05$ for the Alfv\'enic Mach number, very close to the value predicted by Petschek and in agreement with Lyubarsky \cite{2005lyubarsky}, who generalized the Petschek model to the relativistic case and predicted a value of $\mathcal{R}\sim0.1$ for $c_A\rightarrow c$,~\ie, twice the value we used. 
We also measured the angle $\theta$ between the magnetic field and the plane of the shock, that is, the inclination of the magnetic field lines with respect to the current sheet at the slow shock in the exhaust. We find a value of $\theta\sim 0.04 $, compatible with the maximum of the inflow velocity at the X-point, $v_{in} = \tan\theta\simeq\theta$ and again in agreement with \cite{2005lyubarsky}.

\section{Conclusions}
In this work we presented a unified study on the ideal tearing instability in classic, Hall, and relativistic plasmas, carried out by means of 2D fully nonlinear numerical simulations.
Our results confirm that magnetic reconnection via the ideal tearing instability is indeed an efficient mechanism of energy conversion, which is as fast as ideal Alfv\'en timescales in MHD and Hall plasma, and up to light timescales in relativistic environments.

In the frame of MHD and Hall-MHD, after the ideal tearing instability saturated and the nonlinear phase has begun, we observed the onset of secondary reconnection events in newly formed current sheets, thinner of  one order of magnitude than the initial current sheet. These secondary tearing instabilities are stronlgly superalfv\'enic, with reconnection rates $\gamma_{rec}\tau_A\simeq 2.3$, \ie, five time faster than the main instability for the MHD case.
The net result is a much more violent reconnection process and a speed up in the disruption of the current sheet.  
Moreover, numerical simulations performed with increasing values of $\eta_H=d_i/L$ have showed that even though the Hall term is negligible in the linear phase, it considerably affects the secondary reconnection events in the nonlinear phase, by increasing the reconnection rates 
up to about $100\%$ (for $\eta_H=0.0014$) with respect to the pure MHD case and about ten times the reconnection rate of the linear phase. 
The last stage, before the current sheet disrupts, is very turbulent and characterized by enhanced reconnection rates, in agreement with \cite{2004smith}.

Within the relativistic resistive MHD, the tearing instability behaves exactly as in the classical MHD case, provided the Alfv\'en speed is replaced by its relativistic counterpart and the rest mass density $\rho_0$ is replaced by the enhanced inertia term $\rho_0 + 4P_0 +B_0^2$.
Moreover, when rescaled to the light crossing time scale $\tau_c=L/c$,
the linear growth rate of the fastest reconnecting mode in a current sheet of ideal aspect ratio $L/a=S^{1/3}$ is found to be
$\gamma\tau_c \simeq 0.6c_A/c$, indeed very fast and comparable to 1 when the Alfv\'en speed approaches the speed of light.
In the subsequent nonlinear phase we see the merging of the plasmoids and the onset of  secondary reconnection events.
The final stage is dominated by superalfv\'enic Petschek-like jets originating from a single X-point and ending into a huge plasmoid through a fast magnetosonic shock. 
Finally, we retrieve the Petschek reconnection rate $\mathcal{R}\simeq (\ln S)^{-1}$, in agreement with \cite{2005lyubarsky}.
The full evolution of the reconnection process, from the linear phase to the final Petschek state, only last for $t\simeq 10 \tau_c$ when $c_A\rightarrow c$.
All simulations performed show that the evolution of the ideal tearing reconnection is the same independently of the initial conditions and only depends on the Alfv\'en speed $c_A$, at least in the parameter range we explored for the plasma magnetization (up to $\sigma_0=50$) and the plasma beta (down to $\beta_0=0.01$). 

Our results on the ideal tearing have potential applications in space and astrophysical plasmas, \eg, in modeling solar flares and coronal mass ejections, or in improoving our understanding of turbulence in the solar wind. For instance, 
recent numerical simulations retaining kinetic effects \cite{2017franci} assessed the role of reconnection in driving the turbulent cascade at sub-ion scales, through the destabilization of current sheets of thickness $a \simeq d_i$.
Moreover, several theoretical works \cite{2017mallet,2017boldyrev,2017loureiro} strongly suggest that the characteristic spatial and temporal scales introduced by reconnecting current sheets embedded in turbulent environments are crucial in shaping the spectral and statistical properties of turbulence. 
Other astrophysical application in relativistic environments include giant flares observed in the magnetosfere of magnetars \cite{2016elenbaas}, and gamma-ray flares observed, \eg, in the Crab nebula \cite{2014cerutti}.

Although not included in our study, we finally remark that in a full 3D system, different geometrical configurations and intrinsically 3D instabilities, both in MHD and kinetic regimes, may either enhance \cite{2017che} or inhibit \cite{2015gingell} magnetic reconnection. 

\ack
The authors wish to acknowledge valuable exchanges of
ideas with T. Tullio.
EP thanks L. Franci for useful discussions.
SL and LDZ acknowledge support from the PRIN-MIUR project prot. 2015L5EE2Y {\it Multi-scale simulations of high-energy astrophysical plasmas}.
This research was conducted with high performance computing
(HPC) resources provided by the CINECA ISCRA initiative
(grant HP10B2DRR4).

\section*{References}
\bibliography{astronum_bib}

\end{document}